\documentclass[journal=jpccck, manuscript=article, layout=twocolumn]{achemso}

\usepackage{chemformula} 
\usepackage[T1]{fontenc} 
\usepackage{atbegshi}
\usepackage{placeins} 



\let\oldmaketitle\maketitle  
\let\maketitle\relax

\author{Anja Haags}
\affiliation[Juelich]{Peter Grünberg Institut (PGI-3), Forschungszentrum Jülich, 52425 Jülich, Germany}
\alsoaffiliation[JARA]{Jülich Aachen Research Alliance (JARA), Fundamentals of Future Information Technology, 52425 Jülich, Germany}
\alsoaffiliation[AachenIVA]{Experimental Physics IV A, RWTH Aachen University, 52074 Aachen, Germany}

\author{Alexander Reichmann}
\affiliation[Graz]{Institute of Physics, University of Graz, NAWI Graz, 8010 Graz, Austria}
\altaffiliation{Current Address: Chair of Physical Metallurgy, University of Leoben, 8700 Leoben, Austria}

\author{Zilin Ruan}
\affiliation[Marburg]{Department of Chemistry, Marburg University, 35037 Marburg, Germany}

\author{Qitang Fan}
\affiliation[Marburg]{Department of Chemistry, Marburg University, 35037 Marburg, Germany}
\altaffiliation{Current Address: Hefei National Research Center for Physical Sciences at the Micro scale, Synergetic Innovation Center of Quantum Information \& Quantum Physics, and New Cornerstone Science Laboratory, University of Science and Technology of China, Hefei, Anhui 230026, China}

\author{\\ Larissa Egger}
\affiliation[Graz]{Institute of Physics, University of Graz, NAWI Graz, 8010 Graz, Austria}

\author{Hans Kirschner}
\affiliation[PTB]{Physikalisch-Technische Bundesanstalt (PTB), 10587 Berlin, Germany}

\author{Tim Naumann}
\affiliation[Marburg]{Department of Chemistry, Marburg University, 35037 Marburg, Germany}

\author{Simon Werner}
\affiliation[Marburg]{Department of Chemistry, Marburg University, 35037 Marburg, Germany}

\author{\\ Olaf Kleykamp}
\affiliation[Marburg]{Department of Chemistry, Marburg University, 35037 Marburg, Germany}

\author{Jose Martinez-Castro}
\affiliation[Juelich]{Peter Grünberg Institut (PGI-3), Forschungszentrum Jülich, 52425 Jülich, Germany}
\alsoaffiliation[JARA]{Jülich Aachen Research Alliance (JARA), Fundamentals of Future Information Technology, 52425 Jülich, Germany}
\alsoaffiliation[AachenIVA]{Experimental Physics II B, RWTH Aachen University, 52074 Aachen, Germany}

\author{Felix L\"{u}pke}
\affiliation[Juelich]{Peter Grünberg Institut (PGI-3), Forschungszentrum Jülich, 52425 Jülich, Germany}
\alsoaffiliation[JARA]{Jülich Aachen Research Alliance (JARA), Fundamentals of Future Information Technology, 52425 Jülich, Germany}

\author{Fran\c{c}ois C. Bocquet}
\affiliation[Juelich]{Peter Grünberg Institut (PGI-3), Forschungszentrum Jülich, 52425 Jülich, Germany}
\alsoaffiliation[JARA]{Jülich Aachen Research Alliance (JARA), Fundamentals of Future Information Technology, 52425 Jülich, Germany}

\author{Christian Kumpf}
\affiliation[Juelich]{Peter Grünberg Institut (PGI-3), Forschungszentrum Jülich, 52425 Jülich, Germany}
\alsoaffiliation[JARA]{Jülich Aachen Research Alliance (JARA), Fundamentals of Future Information Technology, 52425 Jülich, Germany}
\alsoaffiliation[AachenIVA]{Experimental Physics IV A, RWTH Aachen University, 52074 Aachen, Germany}

\author{Serguei Soubatch}
\affiliation[Juelich]{Peter Grünberg Institut (PGI-3), Forschungszentrum Jülich, 52425 Jülich, Germany}
\alsoaffiliation[JARA]{Jülich Aachen Research Alliance (JARA), Fundamentals of Future Information Technology, 52425 Jülich, Germany}

\author{Alexander Gottwald}
\affiliation[PTB]{Physikalisch-Technische Bundesanstalt (PTB), 10587 Berlin, Germany}

\author{Georg Koller}
\affiliation[Graz]{Institute of Physics, University of Graz, NAWI Graz, 8010 Graz, Austria}

\author{Michael G. Ramsey}
\affiliation[Graz]{Institute of Physics, University of Graz, NAWI Graz, 8010 Graz, Austria}

\author{Mathias Richter}
\affiliation[PTB]{Physikalisch-Technische Bundesanstalt (PTB), 10587 Berlin, Germany}

\author{J\"{o}rg Sundermeyer}
\affiliation[Marburg]{Department of Chemistry, Marburg University, 35037 Marburg, Germany}

\author{Peter Puschnig}
\affiliation[Graz]{Institute of Physics, University of Graz, NAWI Graz, 8010 Graz, Austria}

\author{\\ J. Michael Gottfried}
\affiliation[Marburg]{Department of Chemistry, Marburg University, 35037 Marburg, Germany}

\author{F. Stefan Tautz}
\affiliation[Juelich]{Peter Grünberg Institut (PGI-3), Forschungszentrum Jülich, 52425 Jülich, Germany}
\alsoaffiliation[JARA]{Jülich Aachen Research Alliance (JARA), Fundamentals of Future Information Technology, 52425 Jülich, Germany}
\alsoaffiliation[AachenIVA]{Experimental Physics IV A, RWTH Aachen University, 52074 Aachen, Germany}

\author{Sabine Wenzel}
\email{sabine.wenzel@uni-marburg.de}
\affiliation[Juelich]{Peter Grünberg Institut (PGI-3), Forschungszentrum Jülich, 52425 Jülich, Germany}
\alsoaffiliation[JARA]{Jülich Aachen Research Alliance (JARA), Fundamentals of Future Information Technology, 52425 Jülich, Germany}
\alsoaffiliation[Marburg]{Department of Chemistry, Marburg University, 35037 Marburg, Germany}

\title[]
  {
  Multi-Orbital Charge Transfer into Nonplanar Cycloarenes Revealed with CO-Functionalized Tips
  }
  

\begin{document}


\twocolumn[
\begin{@twocolumnfalse}
\oldmaketitle
\begin{abstract}
On-surface synthesis has allowed for the tuneable preparation of numerous molecular systems with variable properties. Recently, we demonstrated the highly selective synthesis of kekulene (>$99$\%) on Cu(111) and isokekulene ($92$\%) on Cu(110) from the same molecular precursor (Ruan \textit{et al.,} \textit{Angew. Chem. Int. Ed.} \textbf{2025}, e202509932). Scanning tunneling microscopy with CO-functionalized tips can identify the single molecules on the basis of their geometric structure at a low coverage on Cu(110), but it also detects complex features due to electronic contributions close to the Fermi energy. Here, we investigate the origin of these features by simulating STM images based on a weighted sum of multiple molecular orbitals, for which we employ weights based on the calculated molecular-orbital projected density of states. This allows for an experimental confirmation of charge transfer from the surface into multiple formerly unoccupied molecular orbitals for single molecules of kekulene as well as isokekulene in its two nonplanar adsorption configurations. In comparison, the area-integrating photoemission orbital tomography technique confirms the charge transfer as well as the high selectivity for the formation of a full monolayer of mainly isokekulene on Cu(110). Our STM-based approach is applicable to a wide range of adsorbed molecular systems and specifically also suited for strongly interacting surfaces, nonplanar molecules, and such molecules which can only be prepared at extremely low yields.
\end{abstract}
\end{@twocolumnfalse}
]

\section{Introduction}
Molecular systems have a wide variety and range of tunable properties, and therefore great potential for applications in electronics \cite{mathewAdvancesMolecularElectronics2018}, as molecular switches \cite{steenMolecularSwitchingSurfaces2023}, sensors \cite{paolessePorphyrinoidsChemicalSensor2017a}, solar cell materials \cite{songPorphyrinsensitizedSolarCells2018a}, and sustainable single-atom catalysts \cite{huangMetalOrganicFrameworks2020}. A diverse class of carbon-based molecules with variable geometric, electronic, optoelectronic, and magnetic properties have been realized in recent years by on-surface synthesis.\cite{houtsmaAtomicallyPreciseGraphene2021,fanBiphenyleneNetworkNonbenzenoid2021, fanOnSurfaceSynthesisCharacterization2020,haagsKekuleneOnSurfaceSynthesis2020,zhuOnSurfaceSynthesisC1442022,gaoOnsurfaceSynthesisDoubly2023,ruanOnSurfaceSynthesisCharacterization2025} The outcome of on-surface synthesis reactions cannot only be steered by the choice of specialized molecular precursors prepared in solution but also by employing different surface compositions and structures.\cite{caiAtomicallyPreciseBottomup2010,bronnerTrackingRemovingBr2015,hanBottomUpGrapheneNanoribbonFabrication2014,simonovGrapheneNanoribbonsCu1112015,schulzPrecursorGeometryDetermines2017} Most recently, we have reported the highly selective synthesis of kekulene on Cu(111) and its nonplanar isomer isokekulene on Cu(110) by different intramolecular dehydrogenation reactions within the same precursor molecule.\cite{ruanHighlyStructureSelectiveOnSurface}. Single isokekulene molecules on Cu(110) were identified by constant-height scanning tunneling microscopy (STM) using functionalized tips, which is an established technique for imaging the geometric structure of molecules.\cite{temirovNovelMethodAchieving2008,weissImagingPauliRepulsion2010, hapalaMechanismHighresolutionSTM2014,jelinekHighResolutionSPM2017}. However, in this case, the images showed additional complex features due to electronic contributions close to the Fermi energy, which was suggested to be a sign of a strong molecule-surface interaction.\cite{ruanHighlyStructureSelectiveOnSurface} The origin of these features has yet to be investigated in detail. \\ \ \\ 
The electronic structure of adsorbed molecules can be studied with established methods like angle-resolved photoemission spectroscopy (ARPES), especially in combination with density functional theory (DFT) in the form of photoemission orbital tomography (POT).\cite{puschnigReconstructionMolecularOrbital2009, puschnigOrbitalTomographyDeconvoluting2011, wiessnerCompleteDeterminationMolecular2014, luftnerImagingWaveFunctions2014,weissExploringThreedimensionalOrbital2015, zamborliniMultiorbitalChargeTransfer2017, kliuievCombinedOrbitalTomography2019, wallauerTracingOrbitalImages2021, haagsMomentumSpaceImaging2022, yangMomentumselectiveOrbitalHybridisation2022} Recently, POT has been used to study the aromaticity of kekulene synthesized on Cu(111) \cite{haagsKekuleneOnSurfaceSynthesis2020}. Although this technique has so far mostly been applied to planar molecules \cite{puschnigReconstructionMolecularOrbital2009}, there has been recent progress in using it to determine the adsorption geometry of nonplanar molecules \cite{haagSignaturesAtomicCrystal2020,metzgerPlanewaveFinalState2020} as well as to identify nonplanar adsorption geometries of planar molecules \cite{hurdaxLargeDistortionFused2022,janasMetalloporphyrinsOxygenpassivatedIron2023}. However, area-integrating methods like ARPES and POT can only be applied on ordered molecular layers at a significant surface coverage. In contrast, on-surface synthesis, especially when performed by tip manipulation, often leads to low coverages or even single molecules of the desired product, and a number of byproducts or unreacted precursor molecules might be present on the same surface.\cite{gaoOnsurfaceSynthesisDoubly2023, ruanOnSurfaceSynthesisCharacterization2025} \\ \ \\ Scanning tunneling spectroscopy (STS) is well-suited to detect electronic states of single adsorbed molecules. By tuning the bias voltage in the STM to the energetic positions of specific orbitals, images of the real space distribution of the highest occupied and lowest unoccupied molecular orbitals can be recorded.\cite{pascualSeeingMolecularOrbitals2000,martinez-galeraImagingMolecularOrbitals2014a} Other orbitals can be isolated by recording differential conductance (d$I$/d$V$) images at the specific energies.\cite{luSpatiallyMappingSpectral2003,soeDirectObservationMolecular2009,martinez-castroDisentanglingElectronicStructure2022} For nonplanar molecules, the agreement with the theoretical local density of states can be improved by conducting the measurements at constant d$I$/d$V$ instead of constant current or height.\cite{reechtImagingIsodensityContours2017d}. The resolution of both STS spectra and d$I$/d$V$ images can also be improved by tip-functionalization.\cite{martinez-castroDisentanglingElectronicStructure2022}. Specifically, functionalization with CO allows for an additional tunneling channel into \textit{p}-wave orbitals of the tip, which leads to imaging that can be related to the lateral derivative of the local density of electronic states \cite{grossHighResolutionMolecularOrbital2011}. However, even with functionalized tips, specific molecular orbitals might lie too close in energy to distinguish them in single d$I$/d$V$ images. Recently, some of the authors of the present study have demonstrated the deconvolution of orbitals as close as $50$~meV in energy by employing a detailed STS analysis,\cite{martinez-castroDisentanglingElectronicStructure2022} the so-called feature detection algorithm.\cite{sabitovaLateralScatteringPotential2018} However, this approach is based on recording a complete d$I$/d$V$ spectrum at every pixel of an image of the molecule, which is highly time-intensive and requires excellent stability of the sample and the microscope. This is challenging for nonplanar molecules and the energetic separation that can be resolved has a lower limit. Especially multiple almost degenerate orbitals or those broadened strongly due to hybridization with the surface states are challenging to identify with STS.
\begin{figure*}[]
\centering
\includegraphics[width=\linewidth]{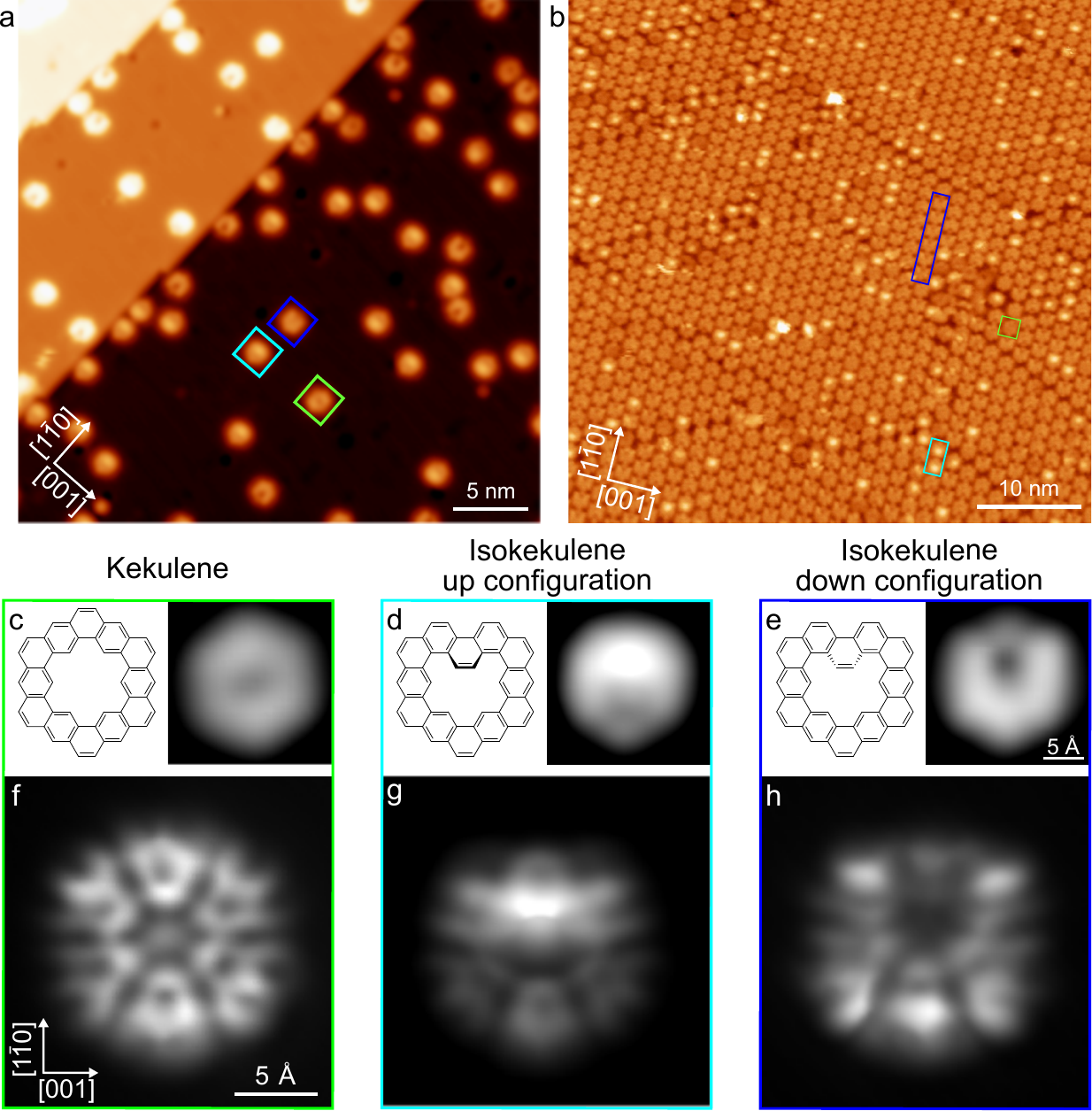}
\caption{Overview STM images of (a) a low coverage and (b) a full monolayer of mainly isokekulene and small amounts of kekulene on Cu(110). (c-e) Small-scale STM images of the surface in (a) showing single molecules of (c) kekulene and isokekulene (d) in the up configuration as well as (e) in the down configuration compared to their respective chemical structures. The STM images were recorded in constant-current mode with (a) $U=50$~mV and $I=20$~pA, (b) $U=0.76$~V and $I=0.18$~nA, (c) $U=-1$~V and $I=150$~pA, (d,e) $U=90$~mV and $I=50$~pA. (f-h) Constant-height STM images recorded with a CO-functionalized tip after stabilization at (f,h) $U=5$~mV or (g) $U=20$~mV and $I=20$~pA above the copper substrate and subsequently increasing the height of the tip by (f) $z=120$~pm, (g) $130$~pm, and (h) $125$~pm.}
\label{STM:samples}
\end{figure*}
\\ \ \\
\indent Here, we present an alternative approach to \FloatBarrier \noindent testing the occupation of multiple orbitals at a specific energy for nonplanar, strongly interacting molecules using only single constant-height STM images recorded with a CO-functionalized tip. In an analogous fashion to POT, the measured STM images are compared to simulations constructed as the weighted sum of theoretical orbitals with the weights determined based on the molecular-orbital projected density of states (MOPDOS) calculated from DFT. The results of this analysis are evaluated in comparison to the results of POT on a full monolayer of the same molecule.
\section{Results and Discussion}
Isokekulene and small amounts of kekulene on Cu(110) are prepared by a combined in-solution and on-surface approach, as reported previously \cite{ruanHighlyStructureSelectiveOnSurface}. Namely, the precursor 1,4,7(2,7)-triphenanthrenacyclononaphane-2,5,8-triene is prepared in solution (details see Ref. \citenum{haagsKekuleneOnSurfaceSynthesis2020}) and vapor-deposited onto the surface at $300$~K. Under annealing to $500$~K, cyclodehydrogenation leads to the respective products. Figure \ref{STM:samples}a and \ref{STM:samples}b shows overview STM images of  the resulting isokekulene molecules (typical examples are marked with cyan and dark blue squares) and small amounts of kekulene molecules (a representative example is marked with a green square) at two different coverages on Cu(110). STM images of single molecules of each species are depicted in Figure
1c to 1e in comparison to the respective chemical structures. Note that isokekulene is nonplanar and adsorbs in two different configurations, denoted as up and down, which differ by the central benzene ring in the molecule's pore pointing upward from or downward toward the surface.
 \\  \ \\
\begin{figure*}[b]
\centering
\includegraphics[width=\linewidth]{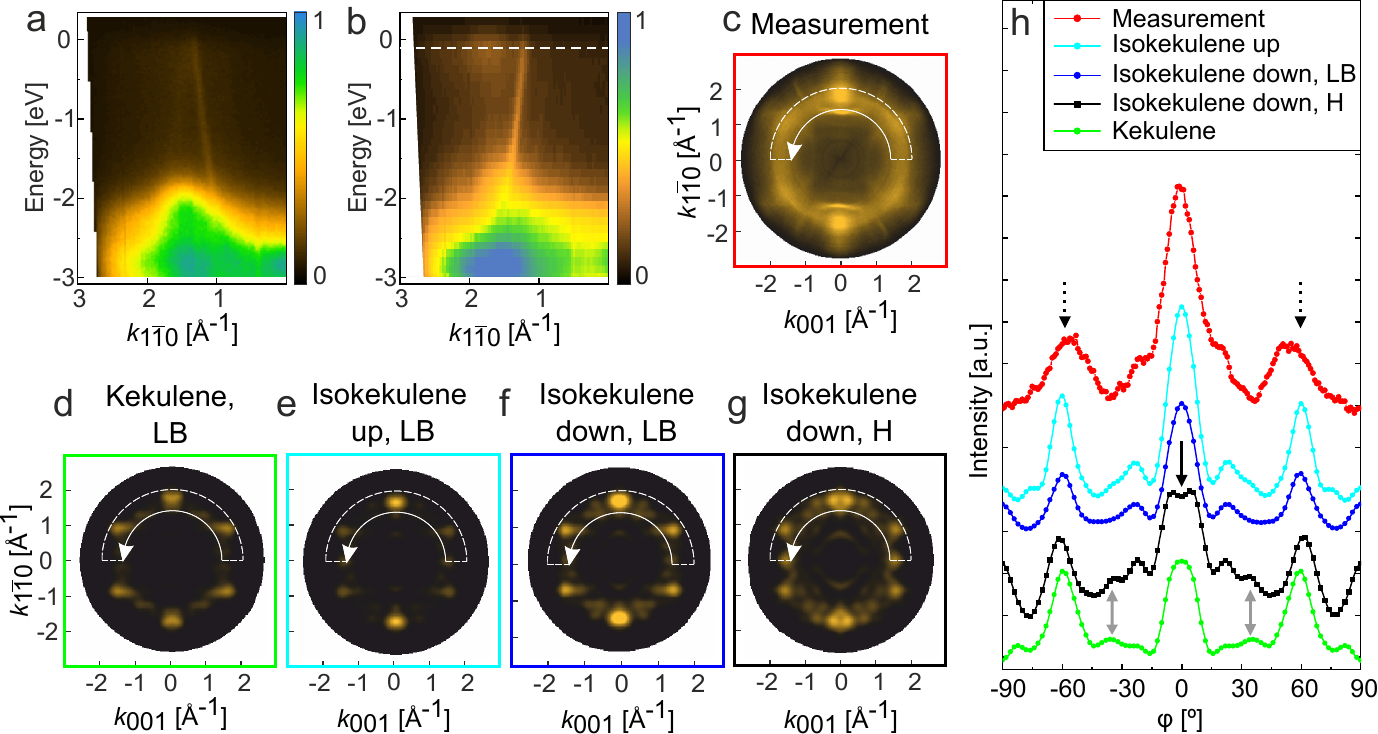}
\caption{(a) Band map measured along the [1$\bar{1}$0]-direction of Cu(111) in an energy range of $3$~eV below the Fermi energy on a monolayer of kekulene and (b) band map measured along the [1$\bar{1}$0]-direction of Cu(110) on a monolayer of (iso)kekulene prepared in the same manner as the surface shown in Figure \ref{STM:samples}b. (c) \textit{k} map measured at $0.11$~eV below the Fermi energy (at the white \textcolor{black}{dashed} line in (b)) and the corresponding simulated \textit{k} maps for (d) kekulene, (e) isokekulene in the up configuration, \textcolor{black}{and (f) isokekulene in the down configuration, all at the long bridge (LB) adsorption site, as well as (g) isokekulene in the down configuration at the hollow (H) adsorption site.} Note that the theoretical \textit{k} maps \textcolor{black}{include contributions of symmetry-equivalent domains that are} indistinguishable by the area-integrating POT. (h) Intensity profiles extracted from the white half circles in the \textit{k} maps. For clarity, the profiles for the different species are shifted along the intensity axis.
}
\label{POT}
\end{figure*} 
 \indent For a more detailed investigation of the density of states around the Fermi energy, the three species were also imaged with a CO-functionalized tip at small bias voltages in constant-height mode 
as shown in Figure \ref{STM:samples}f to \ref{STM:samples}h. These images are in rough qualitative agreement with the images presented in Figure 2i to 2k of Ref. \citenum{ruanHighlyStructureSelectiveOnSurface}, but, here, they were measured at a greater height above the surface (on the order of tens of nanometers higher) in order to reduce any contributions from bond-resolved imaging and therefore remove contributions of molecular geometry to the STM contrast. This is not completely possible for isokekulene in the up configuration (see Figure \ref{STM:samples}g), for which the up-facing central benzene ring still forms a small sharp edge in the contrast, which is indicative of the bond-resolved imaging. When the height was increased even further, the rest of the molecule was not clearly visible anymore. For all three molecules, the presence of a contrast apart from bond-resolved contributions at the low bias voltages used in these measurements suggests that electronic states are present close to the Fermi energy. \\ \ \\
\indent For a detailed investigation of these electronic states, we first applied angle-resolved photoemission spectroscopy, which is an established area-integrating technique complementary to the local STM probe. For comparison, we first discuss an ARPES band map of a monolayer of kekulene on Cu(111) measured along the $[1\bar{1}0]$-direction as depicted in Figure \ref{POT}a. This band map does not show any molecular emissions close to the Fermi energy and is in agreement with the one measured on the same surface in Figure 3a of Ref. \citenum{haagsKekuleneOnSurfaceSynthesis2020}, where a weak molecule-surface interaction was found. In contrast, the band map depicted in Figure  \ref{POT}b, which is measured along the $[1\bar{1}0]$-direction of Cu(110) for a monolayer mainly consisting of isokekulene (as shown in Figure \ref{STM:samples}b), shows clear signatures of molecular emissions just below the Fermi level. \\ \ \\  	\indent
In order to identify the origin of these emissions, we have calculated the electronic structure of kekulene and isokekulene on copper using DFT. These calculations are based on the relaxed adsorption geometries of the three species presented previously \cite{haagsKekuleneOnSurfaceSynthesis2020, ruanHighlyStructureSelectiveOnSurface} and visualized here in Figure \ref{DFT}a to \ref{DFT}e. The lower calculated adsorption height of kekulene on Cu(110)\cite{ruanHighlyStructureSelectiveOnSurface} compared to the same molecule on Cu(111)\cite{haagsKekuleneOnSurfaceSynthesis2020} already suggests a stronger molecule-metal interaction on the more open (110) surface. Additionally, a comparison of the energies presented \FloatBarrier \noindent previously\cite{ruanHighlyStructureSelectiveOnSurface} suggests that kekulene and isokekulene in the up configuration prefer the long bridge site on Cu(110) (see Figure \ref{DFT}b and \ref{DFT}c), whereas the hollow site would be energetically favorable for isokekulene in the down configuration (see Figure \ref{DFT}e). The latter theoretical prediction is in conflict with experimental observations since bond-resolved STM suggests that all three species likely sit in the long bridge site on the low-coverage surface \cite{ruanHighlyStructureSelectiveOnSurface}. The adsorption site in the full monolayer will be discussed below.   \\ \ \\ 
\begin{figure*}[]
\centering
\includegraphics[width=\linewidth]{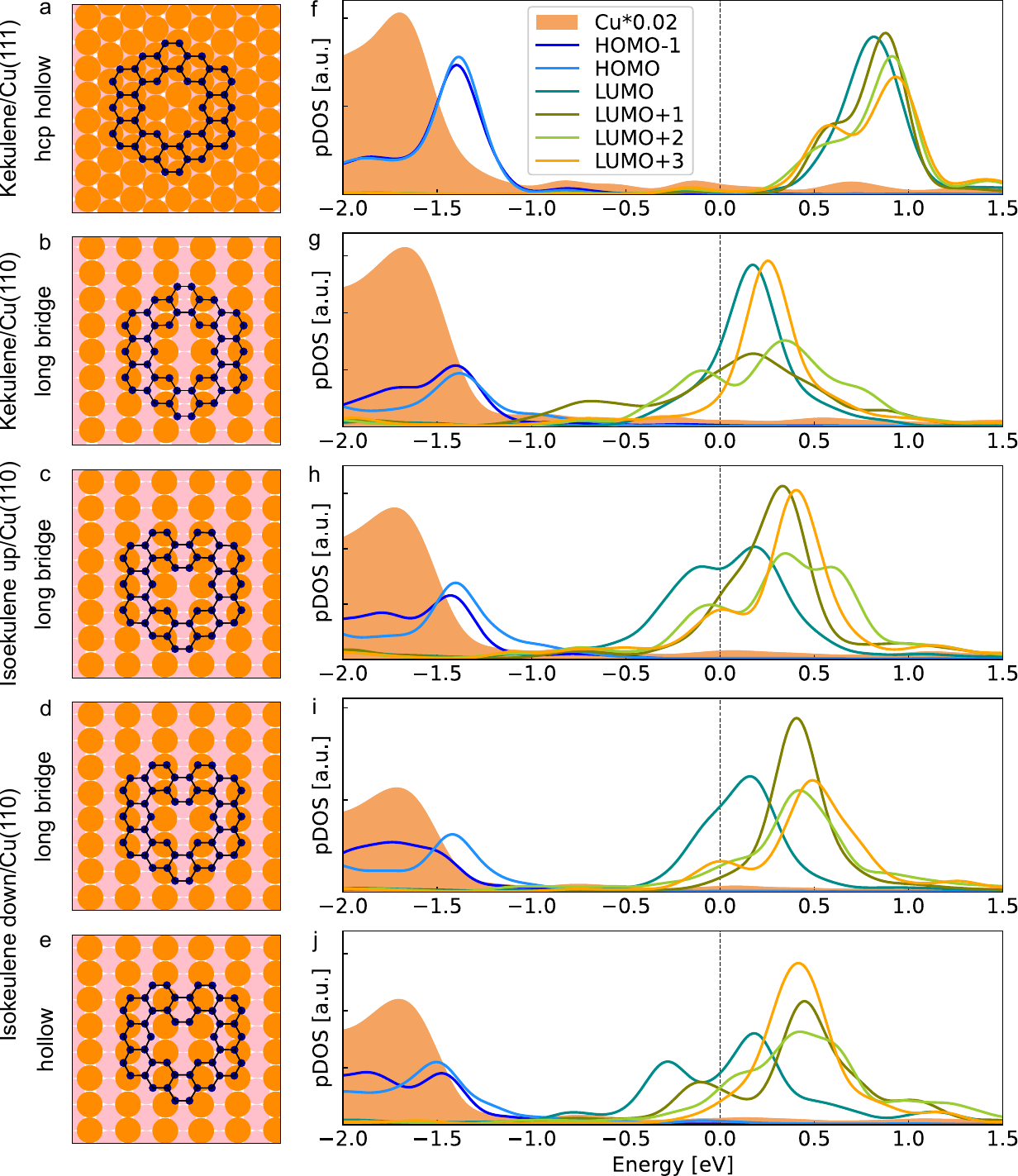}
\caption{(a-e) Visualization of the adsorption of kekulene (a) at the hcp hollow site of Cu(111) and (b) the long bridge site of Cu(110), as well as isokekulene (c) in the up configuration at the long bridge site on Cu(110), and  isokekulene in the down configuration (d) at the long bridge site as well as (e) the hollow site on Cu(110). Copper atoms are orange (top layer) and light pink (second layer), whereas carbon atoms are dark blue. (f-j) Molecular orbital projected density of states (MOPDOS) for HOMO$-$1 to LUMO$+$3 of the corresponding molecular layers including the density of states of copper (the latter scaled by a factor of 0.02). Energies are given relative to the Fermi energy. A Gaussian broadening of 0.1~eV was used.}
\label{DFT}
\end{figure*}
\indent For all adsorption geometries visualized in the left column of Figure \ref{DFT}, the respective MOPDOS was calculated (see Ref. \citenum{luftnerUnderstandingPhotoemissionDistribution2017} and the Methods section for details) and is depicted in the right column of the figure. Figure \ref{DFT}f and \ref{DFT}g show the results for kekulene, comparing the adsorption on the weakly interacting Cu(111) surface to the adsorption on Cu(110). For interpreting the MOPDOS of the adsorbed molecules, it is worth noting that the gas-phase kekulene molecule belongs to the D$_{\text{6h}}$ symmetry group and the orbitals HOMO/HOMO$-$1 (with E$_{\text{1g}}$ symmetry), LUMO/LUMO$+$1 (with E$_{\text{2u}}$ symmetry), as well as LUMO$+$2/LUMO$+$3 (with E$_{\text{1g}}$ symmetry)  are degenerate, respectively. Additionally, the two unoccupied pairs of orbitals are energetically separated by only about $0.1$~eV. As can be seen in Figure \ref{DFT}f, these (near) degeneracies as well as the HOMO-LUMO gap of 2.4 eV for the gas-phase molecule remain almost unaffected upon adsorption of a layer of kekulene on Cu(111) (although this reduces the molecular symmetry group to C$_{\text{3v}}$, in which these degeneracies may generally be lifted). In contrast, significant changes of both the energy level alignment and the hybridization of the molecular and substrate states are observed upon adsorption on Cu(110) (see Figure \ref{DFT}g). Several formerly unoccupied orbitals, ranging from LUMO to LUMO$+$3, appear partially filled, which suggests a charge transfer from the surface to the molecule. As reflected by the broadening of the orbitals in the MOPDOS, we observe a stronger hybridization effect on LUMO$+$1 and LUMO$+$2 compared to LUMO and LUMO$+$3. A similar charge transfer and hybridization as for kekulene on Cu(110) is also found for isokekulene on Cu(110) in the up configuration (see Figure \ref{DFT}h), as well as the down configuration at both adsorption sites which were considered (see Figure \ref{DFT}i and \ref{DFT}j). \\ \ \\
For an experimental confirmation of this predicted charge transfer into multiple molecular orbitals, we additionally applied the ARPES-based POT technique.
For this, we measured the photoemission angular distribution at $0.11$~eV binding energy (marked with the dashed white line in Figure \ref{POT}b) and plotted the corresponding \textit{k} map (or momentum map, calculated from the angular distribution as described in Ref. \citenum{damascelliProbingElectronicStructure2004}) in Figure \ref{POT}c. The pattern shows two bright emissions at ($0.0$, $\pm 1.9$)~\AA$^{-1}$ and two pairs of elongated emission lobes at ($\pm 1.6$,~$\pm 1.0$)~\AA$^{-1}$, which suggests a good orientational order of the organic film. Yet, a diffuse, ring-like intensity at k$_{||}$ $\approx$ 1.6~\AA$^{-1}$ points to a certain degree of disorder. We compare this experimental \textit{k} map with \textit{k} maps simulated for the relaxed geometries and predicted MOPDOS (Figure \ref{DFT}) of kekulene and isokekulene adsorbed on Cu(110), respectively. Using a damped plane wave as final state and following the approach described in Ref. \citenum{luftnerUnderstandingPhotoemissionDistribution2017}, we compute the \textit{k} maps depicted in Figure \ref{POT}d to \ref{POT}g. For kekulene and isokekulene in the up configuration, we used the long bridge site as the configuration found to be most stable in DFT. 
For isokekulene in the down configuration, both the long bridge and the hollow sites were considered. For a detailed comparison, intensity profiles are extracted from all \textit{k} maps. To this end, the intensities in the half circles marked with the thin white lines are integrated over the radial direction from  $1.4~$\AA$^{-1}$ to $2.0~$\AA$^{-1}$ and displayed as a function of the angle (see Figure \ref{POT}h). First, the profile for isokekulene in the down configuration at the hollow site (black curve) has a dip in the middle of the center peak at $\varphi = 0 ^{\circ}$ (indicated by the black arrow) and additional small peaks around $\varphi = \pm 35 ^{\circ}$ (indicated by gray arrows), which is inconsistent with the measured profile (red). Thus, it can be excluded that the full monolayer consists of \FloatBarrier \noindent isokekulene in the down configuration sitting at the hollow site. For kekulene (green profile) the intensities of the center peak at $\varphi = 0 ^{\circ}$ and the peaks at around $\varphi = \pm 60 ^{\circ}$ (indicated by the dotted arrows and present in all profiles) reach the same intensity level again in disagreement with the measurement. In contrast, the profiles of isokekulene at the long bridge site are qualitatively the same for the up and the down configurations (cyan and dark blue profiles) in that both show a smaller intensity at $\varphi = \pm 60 ^{\circ}$ compared to $\varphi = 0 ^{\circ}$. This gives the best agreement with the measured intensity profile, thus proving that the monolayer consists of mainly isokekulene occupying the long bridge site. This is the same site which was found previously for single molecules.\cite{ruanHighlyStructureSelectiveOnSurface} Additionally, we can conclude that the area-integrating POT allows for a clear distinction between the planar kekulene and its nonplanar isomer, which at the macroscopic scale once again confirms the selectivity towards isokekulene that was found at the molecular scale by STM \cite{ruanHighlyStructureSelectiveOnSurface}. Generally, the good agreement between the measured and the simulated \textit{k}-maps for isokekulene in the long bridge site confirms the charge transfer from the surface into multiple formerly unoccupied molecular orbitals as predicted by the calculated MOPDOS.\\

\begin{figure*}[]
\centering
\includegraphics[width=0.9\linewidth]{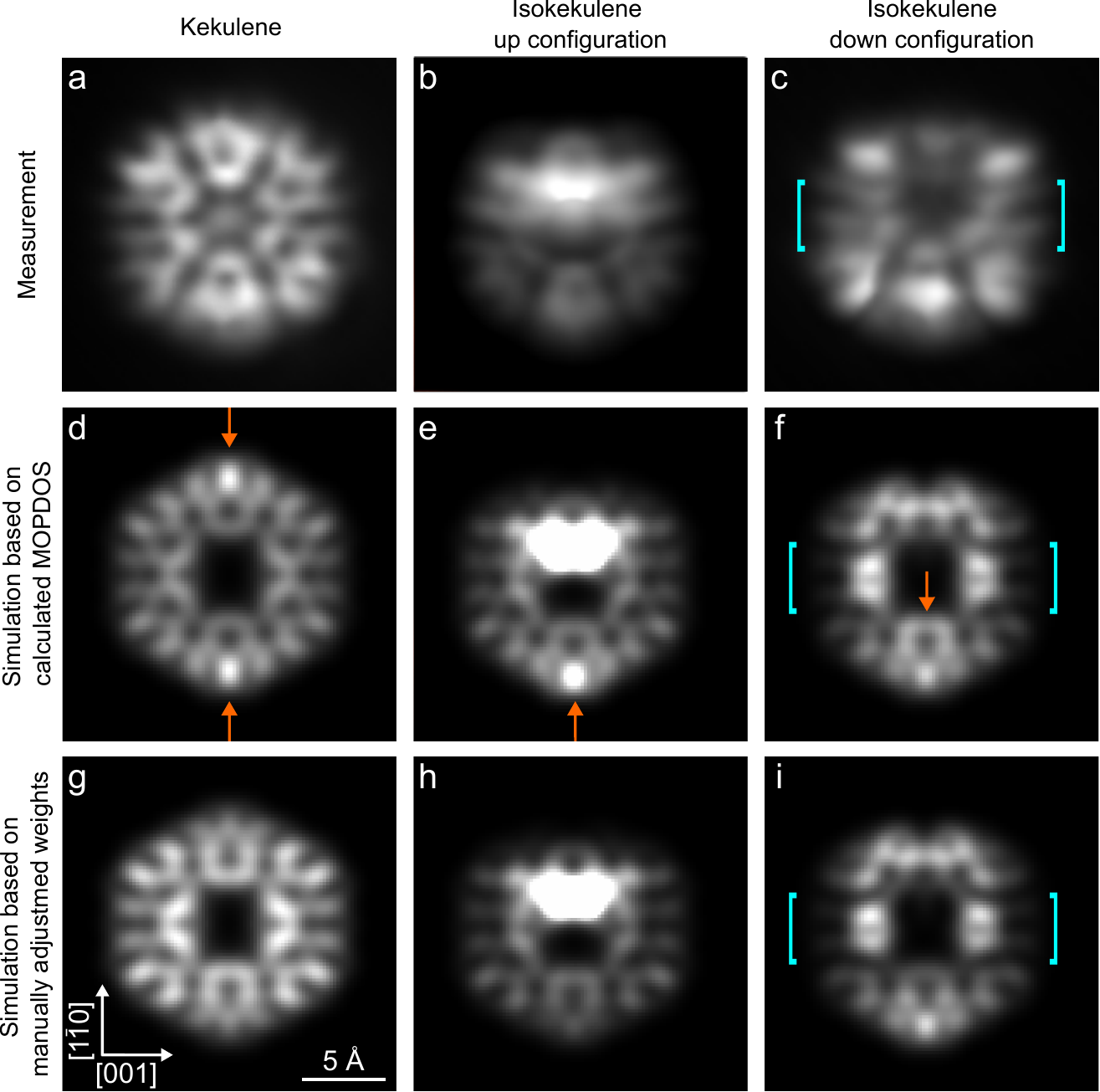}
\caption{STM images of kekulene (left), isokekulene in the up configuration (middle), and isokekulene in the down configuration (right) on Cu(110), recorded in constant-height mode with a CO-functionalized tip (top row), compared to simulated STM images (middle and bottom row).
The simulations employ a \textit{p}-wave tip and are based on a weighted sum of the molecular orbitals LUMO, LUMO$+$1, LUMO$+$2, and LUMO$+$3 in the ratio (d) 36:25:21:14 (e) 33:23:19:18, and (f) 46:8:14:17 (based on the calculated MOPDOS), whereas LUMO$+$1 and LUMO$+$2 are removed from the simulations in (g) and in (h,i), respectively. For all three species the relaxed geometry of the molecules at the long bride site of Cu(110) has been used. To allow for comparison to the measurements a Gaussian broadening was applied to the simulated STM images, which corresponds to a tip radius of 0.4~\AA. Orange arrows and cyan brackets mark specific areas for which disagreements between measurement and simulation are discussed in the text. The STM images were recorded after stabilization at (a,c) $U=5$~mV or (b) $U=20$~mV and $I=20$~pA above the copper substrate and subsequently increasing the height of the tip by (a) $z=120$~pm, (b) $130$~pm, and (c) $125$~pm.}
\label{HR-STM}
\end{figure*}

\indent In the following, we expand the above approach for comparing the measured to the calculated electronic structure from the momentum space to the real space and from a full monolayer to single molecules. For this, we construct simulated STM images based on the theoretical geometries as well as the predicted MOPDOS (Figure \ref{DFT}) and compare them to the measured STM images (Figure \ref{HR-STM}a to \ref{HR-STM}c). Here, we took a standard \textit{s}-wave tip as well as a \textit{p}-wave tip into account. In general, including any \textit{s}-wave contribution worsened the agreement with the measurements. Thus, we can conclude that the CO-decorated tip behaved as a pure \textit{p}-wave tip during our experiment. As described in Ref. \citenum{grossHighResolutionMolecularOrbital2011}, \textit{p}-wave simulations can be realized by calculating the modulus squared of the lateral gradient of the wave function as determined by DFT. To this end, we employ the gas-phase orbitals of the molecules in the relaxed geometries acquired on Cu(110). Figure \ref{SIM_singles} in the Supporting Information shows the resulting images of all single orbitals considered here. As none of these reproduce the measured images in Figure \ref{HR-STM}a to \ref{HR-STM}c, multiple orbitals need to be taken into account simultaneously. For this, the simulated images in Figure \ref{HR-STM}d to \ref{HR-STM}f are constructed from a sum over the contributing orbitals weighted according to the calculated MOPDOS shown above (Figure \ref{DFT}).
As different STM images taken in a bias voltage range of $\pm 50$~mV appear qualitatively similar (data not shown), we approximated the weights of the orbitals by the ones at the Fermi energy for all species. These simulated images reproduce the general appearance of the experimental images, which confirms that there is electron density in the formerly unoccupied orbitals around the Fermi energy.
\\ \ \\  
\indent However, clear deviations can be recognized in these simulated images as marked with the orange arrows in Figure \ref{HR-STM}d to \ref{HR-STM}f. Namely, the contrast is brighter than in the measured images at the top and bottom of the kekulene molecule as well as the bottom end of isokekulene in the up configuration and the lower end of the pore of isokekulene in the down configuration. Owing to the flexibility of the presented approach to simulate the STM images as a weighted sum of single orbitals, we can, however, easily make adjustments as is shown in Figure \ref{HR-STM}g to \ref{HR-STM}i. For all three molecules, the agreement is improved by removing one of the orbitals, the LUMO$+$1 for kekulene and the LUMO$+$2 for isokekulene in both configurations, from each simulation. The MOPDOS (Figure \ref{DFT}) shows that these orbitals are strongly broadened by the hybridization. However, as the numbering of the orbitals stems from their energetic order in the DFT result for the adsorbed state rather than corresponding to specific orbitals in the gas phase, no further conclusions can be drawn from the identity of the orbitals which had to be removed to improve the agreement for the different molecules. Although the shape of orbitals is a rather robust result of DFT, it is known that the calculated orbital energies can vary depending on the chosen exchange-correlation functional and differ from experimental results in some instances \cite{puschnigEnergyOrderingMolecular2017, haagsMomentumSpaceImaging2022}. This is likely the reason for one orbital not being detected for each species here, which examplifies how the detailed comparison to STM measurements can be used to test theoretical predictions of charge transfer into specific orbitals. \\ \ \\
\indent After the removal of the LUMO+2 from the weighted sum, one remaining deviation between the measured and simulated STM images is visible for isokekulene in the down configuration. The measurement in Figure \ref{HR-STM}c shows a rather flat contrast in the center of the molecule in between the cyan brackets overlaid in the image. In comparison, the same area in both corresponding simulations (see Figure \ref{HR-STM}f and \ref{HR-STM}i) shows a stronger contrast, namely a more pronounced completely dark pore in the center surrounded by much brighter features and again a darker contrast at the outside of this area toward the cyan brackets. The agreement in this area could not be improved by any different combination of all four orbitals in the weighted sum. 
Thus, one possible reason for the disagreement is that it does not stem from the shape of the orbitals itself, but rather from the relaxed DFT-calculated geometry that these orbitals are placed upon. Note that the simulated STM image for isokekulene in the down configuration is based on the calculated geometry at the long-bridge site. Although the hollow site is expected to be most stable according to DFT, the long bridge site was previously identified as the adsorption site in high-resolution STM images of single isokekulene molecules in both configurations on Cu(110) \cite{ruanHighlyStructureSelectiveOnSurface}. Figure \ref{SIM_hollow} shows a comparison between simulated STM images employing the geometry at the two different sites in combination with the different weights determined from the calculated MOPDOS at those sites, with as well as without manual adjustment. Comparing those images based on the same MOPDOS (within one column of Figure \ref{SIM_hollow}) to each other as well as those based on the same geometry (within the top and bottom row of the figure, respectively), it appears that the simulated images are more sensitive to the choice of orbitals than to the geometry these are placed upon.
However, the agreement with the experimental image cannot be improved by choosing either the MOPDOS (see Figure \ref{SIM_hollow}d), the geometry (see Figure \ref{SIM_hollow}f), or both (see Figure \ref{SIM_hollow}h) from the hollow site instead of the long bridge site for the simulation. 
Specifically, in all simulated STM images considered, the molecule appears to be more bent than in the actual experimental geometry, which might also be related to the DFT not correctly identifying the energetically favored adsorption site for isokekulene in the down configuration. The calculated geometries\cite{ruanHighlyStructureSelectiveOnSurface} suggest a height difference of roughly $1$~\AA\ and $1.2$~\AA\ between the centers of the highest and the lowest lying carbon atoms in the molecule at the long bride and the hollow sites, respectively. This theoretically predicted height difference is thus likely overestimated compared to the experimental geometry. A similar discrepancy has previously been identified for the case of PTCDA on MgO using POT.\cite{hurdaxLargeDistortionFused2022} While POT does not detect the discrepancy in the present case and does generally not allow for a distinction between the two different adsorption configurations of isokekulene, the detailed analysis of measured and simulated STM images allows to find a mismatch between theory and experiment for just one of the two configurations.
\section{Conclusion}
In conclusion, we have detected significant charge transfer from Cu(110) into multiple formerly unoccupied molecular orbitals of kekulene and isokekulene by comparing single constant-height STM images to simulations based on DFT results. For this, theoretical gas-phase orbitals are placed onto relaxed adsorption geometries and their contributions are weighted according to the calculated MOPDOS. This approach can be adjusted by removing specific orbitals from the weighted sum to test their occupation at the investigated bias voltage, which serves as a direct comparison of the calculated MOPDOS to a measurement. Whereas satisfactory agreement is achieved for kekulene and isokekulene in the up configuration, a discrepancy persists for isokekulene in the down configuration, which suggests that the DFT does not accurately capture the adsorption geometry and the adsorption site of this species.
In contrast, the area-integrating POT technique cannot distinguish between the two different adsorption configurations of isokekulene, but can identify the long bridge adsorption site in the full monolayer of isokekulene in agreement with our previous STM results for single molecules \cite{ruanHighlyStructureSelectiveOnSurface}.
Overall, the presented combination of experimental and theoretical methods can be widely applied to gain a more comprehensive picture of the geometric and electronic structure of adsorbed planar and nonplanar molecules on weakly as well as on strongly interacting surfaces.
\section{Materials and Methods}
The Cu(110) single crystal with a miscut of less than 0.1$^{\circ}$ was purchased from MaTecK GmbH, Germany, and the surface was prepared by cycles of Ar$^{+}$ ion bombardment and annealing to $850$~K. The precursor was synthesized in solution as reported previously \cite{haagsKekuleneOnSurfaceSynthesis2020}. It was deposited onto the copper surface by evaporation either from a home-built Knudsen cell or from a commercial evaporator from Kentax at $500$~K in ultra-high vacuum (UHV).\\ \ \\
\indent STM measurements on the low-coverage surface were performed on a commercial Sigma LT instrument at liquid-helium temperatures. For the measurements presented in Figure \ref{STM:samples}f to \ref{STM:samples}h and Figure~\ref{HR-STM}, the commercial qPlus sensor was functionalized by a CO molecule. The measurements on the monolayer were conducted at liquid nitrogen temperature on a SPECS STM Aarhus 150 STM. 
The images were processed with WSxM \cite{horcasWSXMSoftwareScanning2007} and Gwyddion \cite{necasGwyddionOpensourceSoftware2012}. Moderate filtering (Gaussian smooth, background subtraction) was applied to the image of the full monolayer (Figure \ref{STM:samples}b).
\\ \ \\
\indent As described previously \cite{ruanHighlyStructureSelectiveOnSurface}, the DFT calculations were performed within a repeated-slab approach employing the Vienna ab-initio simulation package (VASP) and using the projector augmented wave method to treat the core electrons \cite{kresseInitioMolecularDynamics1993, kresseEfficiencyAbinitioTotal1996, kresseEfficientIterativeSchemes1996, kresseUltrasoftPseudopotentialsProjector1999}. Five layers of the copper substrate are modeled with the theoretical lattice parameter of $3.55$~\AA\ as obtained by the PBE-GGA functional including van der Waals corrections according to Tkatchenko and Scheffler \cite{tkatchenkoAccurateMolecularVan2009}. For the adsorption of isokekulene on Cu(110), we use the surface unit cell described by the epitaxial matrix 
$\left(\begin{smallmatrix} 4 & 3 \\ 0 & 6 \end{smallmatrix}\right)$ as observed by low-energy electron diffraction \cite{ruanHighlyStructureSelectiveOnSurface}, while for the adsorption of kekulene the somewhat larger  $\left(\begin{smallmatrix} 5 & 0 \\ 0 & 7 \end{smallmatrix}\right)$ overlayer is necessary to properly fit the molecule into the unit cell. For all cases, we consider four adsorptions sites (top, hollow, short bridge, and long bridge with respect to the central void of (iso)kekulene) and perform geometry relaxations including the two topmost layers of the coppers substrate. For the relaxed geometries, we have evaluated the molecular orbital projected density of states (MOPDOS) and simulated photoemission angular distribution maps (\textit{k} maps) as described in Ref. \citenum{luftnerUnderstandingPhotoemissionDistribution2017}.
\\ \ \\
\indent The simulations of the STM images using a \textit{p}-wave CO tip are performed following Ref. \citenum{grossHighResolutionMolecularOrbital2011}. To this end, we computed the molecular orbitals of the LUMO, LUMO$+$1, LUMO$+$2, and LUMO$+$3 for isolated (iso)kekulene molecules frozen in the adsorption geometry. The resulting STM images are then obtained as weighted averages over the orbital contributions, where the weights are taken from the MOPDOS of the adsorbed molecules.
\\ \ \\
\indent ARPES and POT were conducted at the Metrology Light Source insertion device \cite{gottwaldU125InsertionDevice2019} beamline of the Physikalisch-Technische Bundesanstalt (PTB, Germany). The sample was exposed to \textit{p}-polarized ultraviolet light (35~eV photon energy) with an incidence angle of 40$^{\circ}$ with respect to the surface normal, and photoelectrons were collected with the toroidal electron spectrometer \cite{broekmanFirstResultsSecond2005}. For the measurements shown in Figure \ref{POT}, the photoemission intensity in the emission angle range from $0^{\circ}$ (sample normal) to $+85^{\circ}$ was recorded. Momentum maps were recorded by rotating the sample around its normal in 2$^{\circ}$ steps and measuring the photoemission intensity at a constant kinetic energy of the electrons.
\section*{Acknowledgement}
We thank Hendrik Kaser (Physikalisch-Technische Bundesanstalt, Germany) and John Riley (La Trobe University, Australia) for experimental support during the ARPES and POT measurements. We thank Dominik Brandstetter (University of Graz, Austria) for comments on the manuscript. \\ \ \\
This project has received funding by the Deutsche Forschungsgemeinschaft (DFG, German Research Foundation) through the CRC 1083 "Structure and Dynamics of Internal Interfaces" (grant 223848855) and grant GO1812/4-1. Financial support by the European Research Council (ERC) under the European Union’s HORIZON ERC Synergy Grants action via the project Tackling the Cyclacene Challenge (TACY), grant agreement number 101071420-
TACY-ERC-2022-SYG is gratefully acknowledged. This project was supported by the State of Hessen through the LOEWE Focus Group PriOSS, by the European Regional Development Fund (ERDF), and the Recovery Assistance for Cohesion and the Territories of Europe (REACT-EU). \\ \ \\
\indent Z.R. thanks the Alexander von Humboldt Foundation for a Research Fellowship.
Q.F. thanks the financial support from the Innovation Program for Quantum Science and Technology (2021ZD0303302), the CAS Strategic Priority Research Program, Grant No. XDB 0450201, the CAS Project for Young Scientists in Basic Research (YSBR-054), the National Natural Science Foundation of China (No. 22272156), and the New Cornerstone Science Foundation, China. F.L. and F.S.T. gratefully acknowledge funding from the Bavarian Ministry of Economic Affairs, Regional Development and Energy within Bavaria’s High-Tech Agenda Project ”Bausteine für das Quantencomputing auf Basis topologischer Materialien mit experimentellen und theoretischen Ansätzen”. A.R. and P.P. acknowledge support from the Austrian Science Fund (FWF) project I3731 and the Vienna Scientific Cluster (VSC) for providing the computational resources. F.L. acknowledges funding by the Deutsche Forschungsgemeinschaft (DFG, German Research Foundation) through the Emmy Noether Programme (511 561 801) and Germany’s Excellence Strategy - Cluster of Excellence Matter and Light for Quantum Computing (ML4Q) through an Independence Grant. 
\section*{Supporting Information \\ Available}
Supporting Information available on simulated STM images using single orbitals as well as using different combinations of the geometry and MOPDOS of isokekulene in the down configuration at two different adsorption sites.

\bibliography{references}

\setcounter{figure}{0}
\renewcommand{\thefigure}{S\arabic{figure}}

\begin{figure*}
\flushleft\section*{S1. Supporting Information}
\vspace{3cm}
\includegraphics[width=1\linewidth]{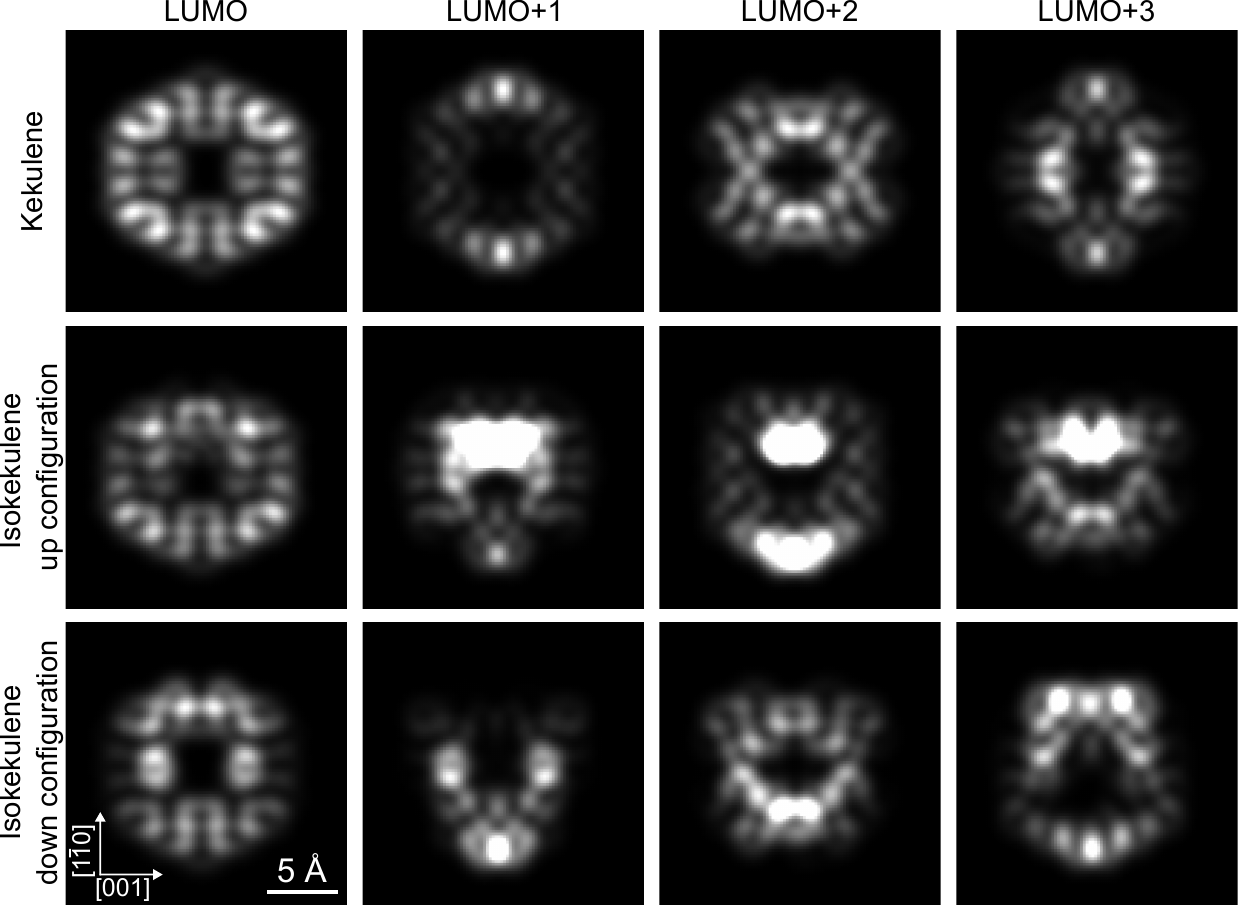}
\caption{Simulated \textit{p}-wave STM images of kekulene (top row), isokekulene in the up configuration (middle row), and in the down configuration (bottom row) based on single orbitals (from left to right: LUMO, LUMO$+$1, LUMO$+$2, and LUMO$+$3). For all three species the relaxed geometry of the molecules at the long bride site of Cu(110) has been used. A Gaussian broadening was applied to the simulated STM images, which corresponds to a tip radius of 0.4~\AA.}
\label{SIM_singles}
\end{figure*}

\begin{figure*}
\includegraphics[width=1\linewidth]{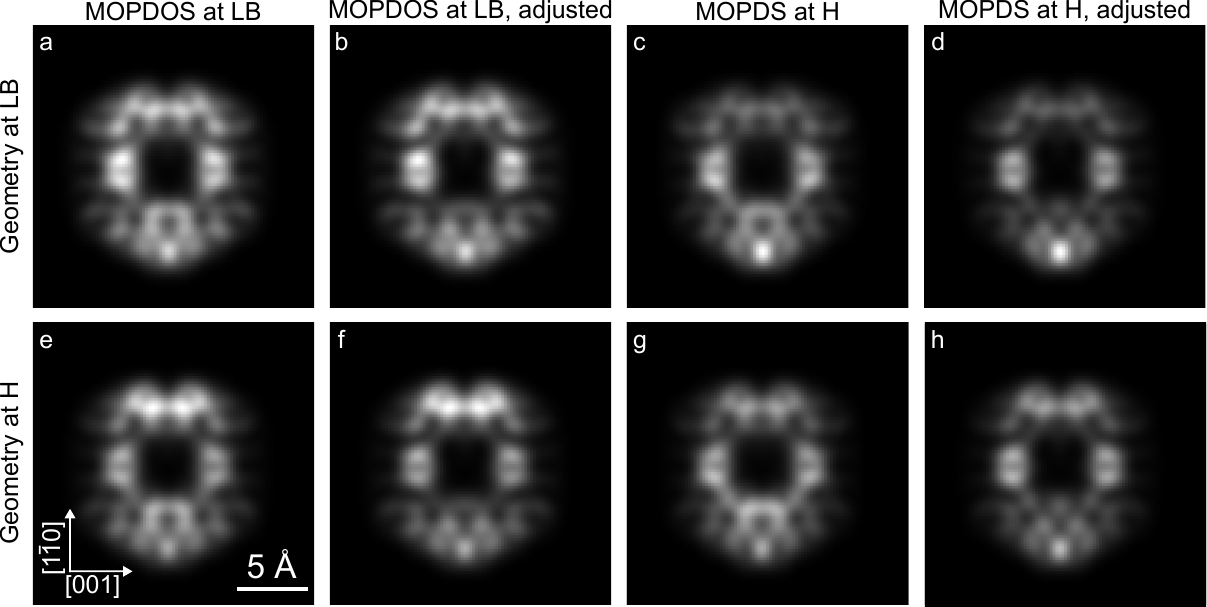}
\caption{Simulated \textit{p}-wave STM images based on the calculated geometries of isokekulene in the down configuration (top row) at the long bridge (LB) site and (bottom row) at the hollow (H) site on Cu(110) employing a weighted sum of the molecular orbitals LUMO, LUMO$+$1, LUMO$+$2, and LUMO$+$3 in the ratio (first column) 46:8:14:17, (second column) 46:8:0:17, (third column) 23:16:16:10, and (forth column) 23:16:0:10. A Gaussian broadening was applied to the simulated STM images, which corresponds to a tip radius of 0.4~\AA.}
\label{SIM_hollow}
\end{figure*}

\end{document}